\documentclass[conference]{IEEEtran}
\usepackage{cite}
\usepackage[OT1]{fontenc}
\usepackage{amsmath,amssymb,amsfonts}
\usepackage{algorithmic}
\usepackage{graphicx}
\usepackage{textcomp}
\usepackage{xcolor}
\usepackage{hyperref}

\usepackage{amsthm}

\def\BibTeX{{\rm B\kern-.05em{\sc i\kern-.025em b}\kern-.08em
    T\kern-.1667em\lower.7ex\hbox{E}\kern-.125emX}}

\begin{document}

\title{Software Supply Chain Security of Web3}

\author{\IEEEauthorblockN{Martin Monperrus}
\IEEEauthorblockA{\textit{KTH Royal Institute of Technology \& \texttt{mab.xyz}}\\
monperrus@kth.se}}

\maketitle

\begin{abstract}
Web3 applications, built on blockchain technology, manage billions of dollars in digital assets through decentralized applications (dApps) and smart contracts. These systems rely on complex, software supply chains that introduce significant security vulnerabilities. This paper examines the software supply chain security challenges unique to the Web3 ecosystem, where traditional Web2 software supply chain problems  intersect with the immutable and high-stakes nature of blockchain technology. We analyze the threat landscape and propose mitigation strategies to strengthen the security posture of Web3 systems.
\end{abstract}


\section{Introduction}

The Web3 ecosystem has rapidly evolved from a technological curiosity into critical financial infrastructure, with decentralized applications managing billions of dollars across DeFi protocols, NFT marketplaces, and decentralized applications \cite{huang2024overview}. Unlike traditional web applications where security failures result in data breaches or service disruptions, vulnerabilities in Web3 systems directly translate to irreversible financial losses, as demonstrated by the \$625 million Ronin Network bridge hack \cite{belenkov2025sok}. Web3 applications is first and foremost software. Any vulnerability in the Web3 software stack leads to catastrophic, irreversible onchain consequences.

Traditional software supply chain security research has primarily focused on Web2 environments \cite{Ladisa2023JourneyTTA,Williams2025ResearchDIA}, examining package repository compromises, dependency attacks, and malicious maintainer scenarios in conventional applications. While these attacks affect availability and confidentiality, Web3 supply chain compromises have different outcomes: they enable direct theft of user funds through transaction parameter manipulation, private key exfiltration, and malicious contract deployment. Recent incidents demonstrate this elevated threat landscape: the 2025 Bybit attack leveraged a compromised frontend JavaScript dependency to modify multi-signature wallet transactions \cite{bybit}. 
Despite the scale of the problem, a systematic analysis of supply chain attack vectors specific to Web3 is missing.

This paper presents a comprehensive analysis of software supply chain security challenges unique to the Web3 ecosystem, examining how blockchain properties of immutability, transparency, and finality create novel attack vectors and amplifies consequences compared to traditional software systems. We develop a threat model that categorizes Web3 supply chain attacks across multiple architectural layers, from blockchain node infrastructure, smart contract dependencies to frontend delivery mechanisms. Our analysis reveals critical vulnerabilities at each layer: compromised RPC endpoints can manipulate blockchain state visible to applications, malicious smart contract library upgrades simultaneously affect all dependent contracts, and frontend dependency poisoning enables transaction parameter modification that bypasses user review.

We make the following contributions:
\begin{itemize}
\item We define Web3 systems and their unique characteristics: immutability, transparency, and finality. These unique characteristics amplify supply chain attack impact compared to traditional software.
\item We present a systematic threat model for Web3 software supply chains, identifying attack vectors across blockchain nodes, smart contract dependencies, frontend libraries, wallet software, development tools, and hosting infrastructure.
\item We propose a defense-in-depth mitigation strategy combining technical controls (dependency verification, reproducible builds, multi-signature deployment), process improvements (specialized audits, incident response), and continuous monitoring for both frontend and on-chain components.
\end{itemize}

\section{Background}
\subsection{Web3 Software Architecture}

\begin{figure*}[t]
\centering
\includegraphics[width=0.7\linewidth]{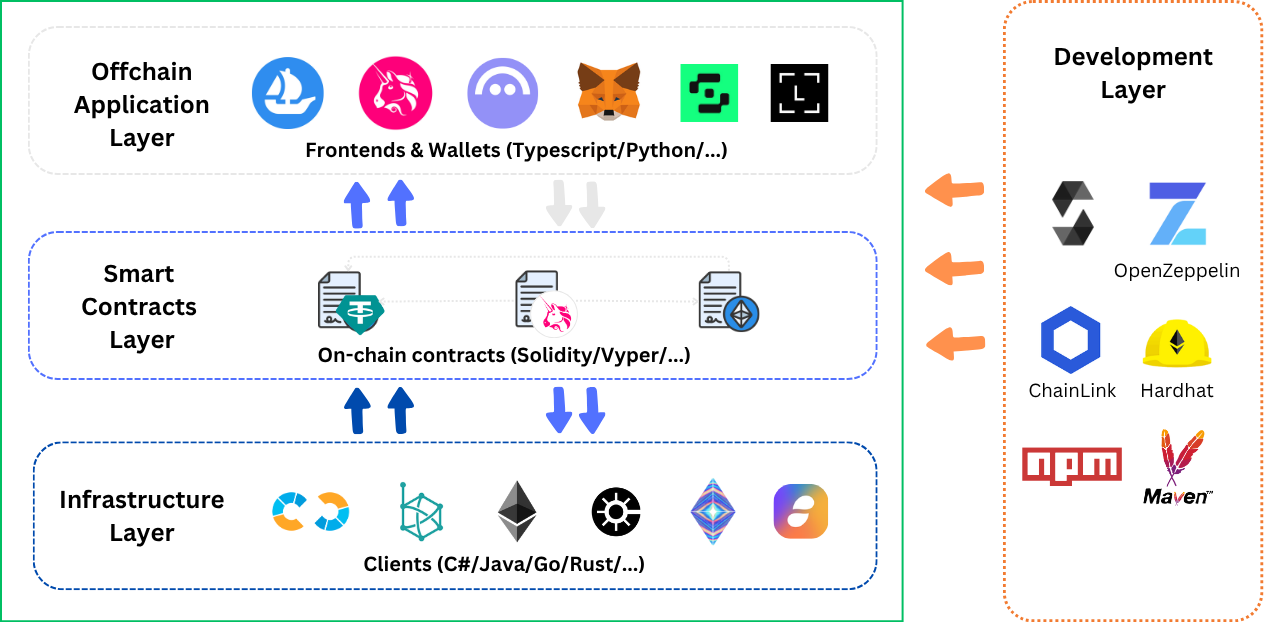}
\caption{A Web3 system has many layers, each of them is subject to software supply chain attacks.}
\label{fig:ecosystem_layers}
\end{figure*}

Web3 applications are different from traditional software systems. A typical Web3 system consists of blockchain nodes, smart contracts, frontend and wallet software.
The smart contracts are deployed on blockchain platforms like Ethereum, Solana, or Polygon. These contracts execute business logic in an immutable, trustless environment. The frontend, usually built with web frameworks, interacts with these contracts through wallet software. Wallets manage cryptographic keys that represent user identity and asset ownership. This architecture can eliminate centralized servers entirely: the backend runs on-chain through smart contracts, while the frontend is hosted on decentralized storage systems like IPFS. Once deployed, these applications can operate autonomously with zero ongoing operational costs, requiring no server maintenance or hosting fees.

Development and deployment workflows in Web3 is also specific. Developers write smart contracts in languages like Solidity or Rust, compile them to bytecode, and deploy to blockchain networks. Testing happens on local nodes, testnets, and eventually mainnet forks before production deployment. Build pipelines are used for both smart contract and frontend development. Deployment requires managing private keys with extremely high value. The Web3 development workflow uniquely combines traditional software engineering with cryptographic operations and irreversible deployments.

\subsection{Software Supply Chain Concepts}

The software supply chain encompasses all components, processes, and actors involved in creating and delivering software from source code to production deployment. In traditional software development, this chain includes package managers (npm, pip, Maven), build systems (Make, Gradle, Webpack), continuous integration platforms (GitHub Actions, Jenkins), and artifact registries (Docker Hub, npm registry). 

The scale of software reuse today is unprecedented. A typical web application depends on hundreds of third-party packages, each with their own dependencies forming a complex dependency tree. Package managers resolve these dependencies, downloading more dependencies, called transitive dependencies. This means developers often execute code they've never reviewed from maintainers they've never vetted. The 2016 left-pad incident, where an 11-line npm package removal broke thousands of projects, demonstrated the fragility of these dependency chains \cite{leftpad}.

Trust in traditional supply chains rely on centralized authority. Package registries verify maintainer identities through email or OAuth. Code signing uses certificate authorities. All, these mechanisms assume registry operators are trustworthy and that compromised credentials or malicious maintainers can be detected \cite{Ohm2020BackstabbersKCA}. The 2018 event-stream attack, where a malicious maintainer added cryptocurrency-stealing code to a popular npm package, revealed the potential of supply chain attacks on crypto \cite{event-stream}.

Note that this paper focuses on software supply chain security in Web3 systems. We do not cover how traditional software supply chain security can be improved with Web3 technology \cite{Zhu2024ESPRAEA,Sommer2023AMAA}.

\section{Threat Model of Web3 Systems}

The unique characteristics of Web3 applications create distinct security considerations compared to traditional software systems. Understanding the threat landscape requires examining how blockchain properties interact with software supply chain vulnerabilities.

\subsection{Unique Properties of Web3 Systems}

The \textbf{immutability} of blockchain means deployed smart contracts cannot be patched or updated without complex upgrade mechanisms, making vulnerabilities permanent without careful planning. Unlike traditional web applications where bugs can be hotfixed within minutes, smart contract vulnerabilities require careful migration strategies, proxy patterns, or complete redeployment with state migration.

The \textbf{transparent} nature of blockchain exposes all contract code and state publicly, allowing attackers to study systems before exploiting them. Every transaction, every balance, and every line of deployed bytecode is visible to potential adversaries. All notable smart contracts are ``verified'', meaning their source code is publicly available alongside the deployed bytecode. This transparency eliminates "security through obscurity" entirely and requires that systems be secure by design rather than by information hiding. 

Onchain operations are \textbf{final} and irreversible. Once a transaction is confirmed, it cannot be undone or rolled back. This finality means that any exploit leading to fund transfers or state changes is permanent, with no recourse for victims. Unlike traditional systems where transactions can be reversed through customer support or legal action, blockchain transactions are final once mined into a block.
This creates unprecedented incentives for attackers, where a single vulnerability can result in the irreversible loss of hundreds of millions of dollars within seconds.

Finally, the \textbf{pseudonymous nature} of blockchain interactions makes attribution difficult, providing attackers with operational security advantages. Attackers can interact with smart contracts through fresh addresses with no connection to their real identity, withdraw funds through mixing services, and operate across jurisdictions with minimal law enforcement risk.

\subsection{Adversary Capabilities and Motivations}

Web3 systems face threats from multiple adversary classes, some of them being extremely powerful. Nation-state actors target Web3 infrastructure to fund covert operations \cite{northkorea}. Organized cybercriminal groups are attracted by the direct monetization potential and difficulty of attribution. Individual opportunistic attackers can exploit discovered vulnerabilities for financial gain. Even legitimate competitors may engage in adversarial behavior within protocol design rules.

Financial motivation dominates the Web3 threat landscape. The immediate liquidity of cryptographic assets through decentralized exchanges means attackers can extract and launder stolen funds within minutes. This tight attack-to-monetization loop creates one of the strongest incentives ever for discovering and exploiting vulnerabilities before defenders can respond.


\section{Software Supply Chain Attack Vectors in Web3}

Web3 applications depend on complex software supply chains spanning smart contract libraries, frontend frameworks, development tools, and deployment infrastructure. Each component in this chain represents a potential attack vector. This section examines the primary attack vectors targeting Web3 software supply chains.

\subsection{Blockchain Nodes}

Blockchain nodes serve as the foundational infrastructure for Web3 applications, providing access to blockchain state and enabling transaction submission. Most Web3 applications and developers rely on third-party node providers like Infura, Alchemy, or Dwellir rather than running their own full nodes due to the significant computational and storage requirements. This dependency creates a critical supply chain vulnerability where compromised or malicious node infrastructure can manipulate application behavior without modifying any application code.

\textbf{RPC Endpoint Attack}
Remote Procedure Call (RPC) endpoints represent the primary interface through which Web3 applications communicate with blockchain networks, making them a critical attack vector in the supply chain. When a dApp queries blockchain state or submits transactions, it sends JSON-RPC requests to node providers that can potentially manipulate the responses. A compromised RPC endpoint could return falsified blockchain data, such as incorrect account balances, fabricated transaction histories, or manipulated smart contract state, leading applications to make decisions based on fraudulent information. An attacker controlling a widely-used RPC endpoint could simultaneously compromise thousands of dApps relying on that infrastructure. This attack vector is particularly dangerous because it is invisible to both end users and developers: applications seem to function normally while receiving manipulated data, and transactions appear to be submitted normally.

\textbf{Dependency Attack on Node Software}
Blockchain node software itself relies on extensive dependency chains that present supply chain attack opportunities. Full node implementations like Geth (Ethereum), Solana validators, or Bitcoin Core depend on hundreds of libraries for networking, cryptography, database management, and consensus logic. A dependency in node software can be compromised through various normal supply chain attack vectors: a legitimate package maintainer account takeover, a typosquatting attack with similarly-named packages, or a compromised build system \cite{Andersson1908608}. Once injected, malicious code in node dependencies could manipulate consensus participation, selectively relay or withhold transactions, return falsified blockchain state to querying applications, or exfiltrate sensitive operational data about network topology and validator behavior. 

\subsection{Web3 Dependencies}

Web3 applications depend on multiple layers of software components, each presenting potential attack vectors in the supply chain \cite{Ma2024DecodingWIA}.

\subsubsection{Smart Contract Dependencies}

Smart contract development relies on smart contract libraries. For example, the OpenZeppelin Contracts library, used in thousands of projects, is a critical dependency where a compromise could affect numerous deployed contracts. Those dependencies are typically managed through npm, introducing the same risks seen in traditional software supply chains.

\textbf{Pre-deployment Attack.} Smart contract libraries imported or copied into projects during development create a pre-deployment attack window. If a malicious version of OpenZeppelin Contracts is installed through npm, the compromised code becomes part of the contract source before compilation and deployment. An attacker who compromises a smart contract library package can inject subtle vulnerabilities -- such as backdoored access control checks, hidden fund drainage mechanisms, or logic bombs triggered by specific conditions -- that survive compilation and remain undetected through standard audit processes. The immutability of deployed contracts means these pre-deployment compromises become permanent vulnerabilities that cannot be patched without complete contract redeployment and user migration.

\textbf{Smart Contract Library Attack.}
Smart contracts can also depend on on-chain libraries deployed as separate contracts. Rather than copying library code into each contract that uses it, developers can deploy libraries once and reference them later one. This reduces deployment costs and allows multiple contracts to share common functionality. For example, complex mathematical operations, signature verification routines, or shared business logic are deployed as library contracts.
Upgradeable on-chain libraries represent a critical attack vector \cite{crystal-clear}. If an upgradeable library is compromised or maliciously upgraded, all contracts depending on it become vulnerable simultaneously. An attacker who gains control of a library's upgrade mechanism can inject malicious code that affects every dependent contract in a single transaction. This creates a many-to-one trust relationship where hundreds of contracts might depend on a single upgradeable library administrator. The 2017 Parity Multisig wallet hack, where a vulnerability in a shared library contract led to \$150 million in frozen funds, demonstrated the catastrophic impact of on-chain library vulnerabilities \cite{parity}. Unlike web2 dependencies (eg npm) which are copied into each deployment, on-chain library attacks can affect all dependent contracts through a single malicious upgrade.

\subsubsection{Frontend Dependencies}

Web3 systems are used through frontend applications that interact with smart contracts. In most of the cases, these frontends are Web applications, built with JavaScript frameworks like React or Vue.js and rely on package managers like npm or yarn to manage dependencies.
Specific blockchain interaction libraries such as ethers.js, web3.js, and viem mediate all communication between applications and blockchain networks. 

These frontend dependencies introduce significant attack surface. Malicious packages or compromised versions can modify transaction parameters before signing, change recipient addresses, exfiltrate private keys from browser storage, or inject phishing overlays that deceive users. An attacker who compromises a popular frontend library could simultaneously affect thousands of Web3 dApps that depend on it. Since frontend code executes with access to wallet APIs and sensitive user interactions, even transitive dependencies deep in the dependency tree can access cryptographic operations and modify transactions. The 2022 compromise of the ua-parser-js npm package \cite{ua-parser}, which was downloaded millions of times per week, demonstrated how widely-used libraries can be weaponized to deliver malware to countless downstream applications.

\textbf{Transaction Manipulation Attack}
A particularly insidious attack vector specific to Web3 frontends is the manipulation of transaction parameters before they are presented to users for signing. When a user initiates a transaction through a dApp interface, the frontend constructs transaction objects containing parameters such as the recipient address, token amount, gas limits, and contract function calls. A compromised dependency in the frontend stack can intercept these transaction objects and modify critical parameters such as changing the recipient address to an attacker-controlled wallet, increasing the token approval amount to the maximum uint256 value, or substituting benign contract calls with malicious ones. Users typically do not understand transaction details they sign in wallet interfaces. The user's signature cryptographically authorizes the malicious transaction, making it indistinguishable from legitimate operations on-chain. This attack is particularly effective because it exploits the trust boundary between the dApp frontend and the wallet: users trust that the dApp correctly constructs transactions matching their intent, while wallets trust the frontend to build the right transaction. 

This attack vector was demonstrated in the 2025 Bybit exchange attack \cite{bybit}, where a compromised Safe dApp frontend Javascript dependency modified transaction parameters before user signing. The malicious code altered a multi-signature wallet transaction, resulting in stealing \$1.5 billion in assets. Recall that the transaction appeared legitimate in the wallet interface and was signed by multiple authorized parties.

\subsection{Wallet Software}

Cryptocurrency wallets represent critical supply chain components that manage private keys and sign transactions \cite{houy2023security}. Compromised wallet software can sign malicious transactions, exfiltrate seed phrases, or manipulate displayed transaction data to deceive users.

\textbf{Wallet Attack} Wallet software itself contains extensive dependency trees that present attractive targets for supply chain attacks \cite{raphinathesis}. Popular wallets like MetaMask, Ledger Live, and Trust Wallet are built using thousands of packages, each representing a potential entry point for malicious code. A typical wallet application depends on cryptographic libraries (elliptic curve operations, hash functions, random number generation), blockchain interaction libraries (JSON-RPC clients, transaction construction utilities), UI frameworks (React, Electron), and numerous utility packages for data parsing, encoding, and validation. Each of these dependencies, including deeply nested transitive dependencies, executes with the same privileges as the wallet itself, meaning access to private key material, seed phrases, and transaction signing operations. An attacker who compromises any package in the dependency tree can potentially exfiltrate keys, modify transactions before signing, or tamper with transaction approval dialogs. The 2023 Ledger Connect Kit compromise \cite{ledger_security_2023} demonstrated this risk when attackers gained control of a library used by Ledger's browser extension, allowing them to inject malicious code that prompted users to approve transactions draining their funds. The high-value target nature of wallets, combined with the trust users place in them and their extensive attack surface through dependencies, makes wallet supply chains a prime focus for sophisticated attackers.

\begin{table*}[t]
\centering
\caption{Web3 Software Supply Chain Attack Vectors}
\label{tab:attack-vectors}
\begin{tabular}{|p{2.5cm}|p{3.5cm}|p{8.5cm}|}
\hline
\textbf{Layer} & \textbf{Attack Vector} & \textbf{Attack Mechanism} \\
\hline
Blockchain Nodes & RPC Endpoint Attack & Compromised node providers return falsified blockchain data or manipulate transaction submission \\
\hline
Blockchain Nodes & Node Software Dependency & Malicious dependencies in node implementations manipulate consensus or transaction ordering \\
\hline
Smart Contract & Pre-deployment Library Attack & Malicious versions of contract libraries inject vulnerabilities during development \\
\hline
Smart Contract & On-chain Library Attack & Compromised upgradeable library affects all dependent contracts simultaneously \\

\hline
Development Tools & Compiler Attack & Malicious compiler injects backdoors during smart contract compilation \\
\hline
Development Tools & IDE Extension Attack & Compromised IDE extensions exfiltrate blockchain private secrets \\
\hline
Development Tools & Framework Attack & Compromised development frameworks modify deployment scripts or contracts \\
\hline
Development Tools & Bundler/Packager Attack & Malicious build tool plugins inject code during frontend bundling \\
\hline
Development Tools & Build Pipeline Attack & Compromised CI/CD systems inject malicious code during automated builds (smart contract or frontend) \\

\hline
Frontend & Frontend Dependency Attack & Compromised frontend libraries modify application behavior \\
\hline
Frontend & Transaction Manipulation & Malicious frontend code modifies transaction parameters before user signing \\
\hline
Frontend & Domain Attack & DNS hijacking redirects users to malicious clones of legitimate dApps \\
\hline
Frontend & Hosting Provider Compromise & Attackers modify deployed frontend artifacts directly \\

\hline
Wallet Software & Transaction Manipulation & Compromised wallet modifies transaction parameters before user signing \\
\hline
Wallet Software & Wallet Dependency Attack & Compromised dependencies in wallet applications access private keys \\

\hline
Infrastructure & Off-chain Voting Attack & Compromised governance platforms manipulate DAO proposals or voting results \\
\hline
Infrastructure & Oracle Software Attack & Compromised oracle software or feed software relay false information to smart contracts \\
\hline
\end{tabular}
\end{table*}

\subsection{Web3 Development Tools and Build Systems}

Compilers, linters, and development plugins used for engineering Web3 systems execute with full access to project source code and developer credentials. 

\textbf{Smart-Contract Compiler Attack} 
A compromised Solidity compiler represents a particularly insidious attack vector in smart contract development because it exploits the compilation process itself \cite{thompson1984reflections}. Since developers and auditors typically review only the Solidity source code, a malicious compiler plugin could silently inject vulnerabilities, backdoors, or unauthorized logic during the compilation phase. For example, it could modify access control checks to allow specific addresses to bypass restrictions, insert hidden transfer functions that siphon funds to attacker-controlled wallets, or alter critical business logic in subtle ways that don't manifest until specific conditions are met. This attack is especially dangerous because the deployed bytecode on the blockchain would differ from what the source code suggests, making it nearly impossible to detect through traditional code audits. This attack highlights the critical importance of using verified compiler toolchains, checking bytecode hashes against known-good versions, and employing reproducible builds for smart contracts, a technique further discussed below.

\textbf{IDE attack} Integrated Development Environments and their extension ecosystems present a significant attack vector in Web3 development due to their privileged access to developer machines. In Web3 development, this creates severe risks because developers frequently store deployment private keys, API keys for blockchain node providers, and other sensitive credentials in environment variables, configuration files (.env files), or even temporarily in clipboard memory during development workflows. A malicious or compromised IDE extension marketed for Solidity syntax highlighting, smart contract debugging, or blockchain interaction could silently exfiltrate these high-value secrets to attacker-controlled servers. The extension malicious code into smart contracts or frontend during editing. The 2025 incident where Visual Studio Code extensions were found to steal cryptocurrency demonstrates this attack vector \cite{rescana_malicious_2025}.

\textbf{Web3 Framework Attack} 
Web3 development frameworks like Hardhat, Truffle, and Foundry serve as comprehensive orchestration layers that manage nearly every aspect of smart contract development, testing, and deployment. These frameworks handle critical security-sensitive operations including private key management for deployment, compilation of smart contracts into bytecode, interaction with blockchain networks through RPC providers, automated testing and verification, and deployment script execution. A compromised framework is a perfect target for injecting malicious code during compilation, exfiltrate deployment private keys that control high-value contracts, or modify deployment scripts to deploy backdoored versions of contracts. These frameworks are centralized:  a single package controls the entire development lifecycle.
This creates a critical single point of failure: compromising Hardhat's npm package could simultaneously affect thousands of projects, potentially leading to coordinated attacks across multiple protocols. The trust developers place in these frameworks, combined with their privileged access to keys, source code, and deployment processes, makes them extremely high-value targets for supply chain attacks with potentially catastrophic ecosystem-wide impact.

\textbf{Packager Attack} Build systems and bundlers process all frontend code before deployment, creating a critical trust boundary where source code is transformed into production artifacts. Tools like Webpack, Rollup, Vite, and Parcel orchestrate complex transformation pipelines involving hundreds of plugins, loaders, and transformers. Each of these components executes with full access to the application's source code and can modify it arbitrarily during the build process.  For example, a malicious bundler plugin could inject code that monitors for high-value transactions and modifies recipient addresses, add logic that exfiltrates user credentials to remote servers. In Web3 contexts, this is particularly dangerous because build tools can modify critical frontend code that interacts with wallets and constructs transactions.

\textbf{Build Pipeline Attack}
A build pipeline attack targets the CI/CD (Continuous Integration/Continuous Deployment) pipeline or build servers of Web3 servers. Attackers may compromise them to inject malicious code during compilation or manipulate dependency management systems to substitute legitimate libraries with compromised versions. Frontend compromises through build pipelines is a traditional software supply chain attack \cite{Ladisa2023JourneyTTA}. In Web3 contexts, these attacks can also target the deployment of smart contracts, because some smart contracts are deployed with CI/CD. Compromised smart contract deployments are immutable once deployed, making build pipeline integrity critical: a single compromised build can result in permanent, irreversible vulnerabilities on the blockchain. 
Also, deployment keys used in CI/CD pipelines may have high privileges, making them attractive targets for attackers.

\subsection{Frontend Delivery \& Hosting}

\textbf{Domain Attack.}
DNS hijacking represents a critical attack vector where adversaries gain control of a dApp's domain name system records to redirect users to malicious infrastructure. Attackers can compromise DNS through multiple vectors: BGP hijacking to reroute DNS queries, social engineering attacks against domain registrars to transfer ownership, or exploitation of vulnerabilities in DNS providers. Once DNS control is obtained, attackers point the legitimate domain (such as app.uniswap.org) to attacker-controlled servers hosting pixel-perfect clones of the authentic dApp interface. Users who navigate to the correct URL have no indication of compromise: the browser displays the expected domain. When users connect their wallets and sign transactions through this malicious interface, they authorize fund transfers to attacker-controlled addresses. This attack bypasses all frontend code integrity mechanisms because the malicious code never enters the legitimate build pipeline; instead, the entire application is substituted at the DNS layer.

\textbf{Hosting Provider Compromise.}
Web3 frontends are typically deployed as static sites on hosting platforms like Vercel, Netlify, AWS S3, or IPFS gateways. Compromise of hosting infrastructure allows attackers to modify deployed artifacts directly without touching the source code or build pipeline. If an attacker obtains credentials for a project's hosting account -- through leaked API keys, compromised CI/CD secrets, or social engineering against platform accounts -- they can inject malicious JavaScript directly into production files. This attack vector was demonstrated in the 2021 BadgerDAO incident \cite{badgerdao} where attackers compromised Cloudflare API credentials and injected malicious scripts, resulting in \$120 million in losses. Unlike dependency poisoning attacks that must survive code review and build processes, hosting compromises modify the final served artifacts after all security checks have completed.

\subsection{Other Web3 Attack Vectors}

\textbf{Off-Chain Voting Platform Attack}
Many DAOs use off-chain voting platforms like Snapshot to conduct governance. These software platforms are a critical part of the ``governance supply chain.'' A compromised Snapshot space, or a vulnerability in the Snapshot frontend itself, could be used to present a malicious proposal as a legitimate one, tricking token holders and delegates into voting for an outcome they don't intend. An attacker who gains control of a DAO's Snapshot space configuration could modify proposal text after voting begins, display falsified voting results to create manufactured consensus, or inject malicious links disguised as legitimate documentation. The off-chain nature of these platforms means votes don't benefit from blockchain immutability and traceability. When DAOs use offchain channels to inform on-chain execution, a compromised voting platform becomes a vector for executing unauthorized governance onchain actions with apparent community approval.

\textbf{Oracle Software Attack}
Oracle software bridges off-chain data to blockchain networks, enabling smart contracts to access real-world information such as asset prices, weather data, or sports results. These systems present a critical supply chain vulnerability. An attacker who compromises oracle node software through dependency poisoning, infrastructure exploits, or malicious updates, can manipulate the data reported to smart contracts. Even more insidious is targeting the upstream data sources themselves: by feeding false information to legitimate APIs that oracles consume (such as exchange price feeds or weather stations), attackers can cause oracle networks to faithfully report malicious data that appears cryptographically valid. Since oracles only verify data integrity, not accuracy, corrupted source data propagates directly on-chain. This enables devastating attacks at the smart contract level: manipulated price feeds triggering mass liquidations in DeFi protocols, falsified sports scores releasing fraudulent prediction market payouts, or fabricated weather data activating insurance claims. 

\textbf{Vibe Coding Attack}
The rise of AI-assisted coding tools like GitHub Copilot, ChatGPT, and specialized Web3 development assistants introduces a novel supply chain attack vector where malicious code can be injected through poisoned training data. When developers use these tools to "vibe code" -- rapidly generating boilerplate smart contracts, frontend integration logic, or wallet interaction code -- the AI models may suggest vulnerable patterns that were deliberately planted in open-source repositories used for training. An attacker who contributes subtly backdoored code to popular repositories can influence the model's learned patterns, causing it to suggest similar vulnerabilities to developers across the ecosystem. For smart contracts, this might manifest as flawed access control patterns or integer overflow conditions in seemingly innocuous arithmetic. For frontend code, poisoned training data could lead AI assistants to suggest transaction construction patterns that are vulnerable to parameter manipulation, insecure private key handling in development environments, or wallet integration code with subtle signature verification bypasses. A real-world incident in 2024 demonstrated this attack vector when a user lost \$2,500 after deploying AI-generated Solana meme token bot code that contained a backdoor to steal private keys \cite{vibecoding}. The malicious code originated from poisoned repositories on Github that specifically targeted ChatGPT's training data. This attack is particularly insidious because developers trust AI-generated code as being based on common patterns from legitimate open-source projects. In reality, training corpora have already been deliberately poisoned at scale with malicious examples.

\subsection{Summary}

Table \ref{tab:attack-vectors} summarizes the attack vectors discussed in this section, categorizing them by layer and describing the attack mechanism.

\section{Mitigation Strategies}

We recommend a defense-in-depth approach combining technical controls, process improvements, and continuous monitoring to mitigate supply chain risks in Web3 systems, build on traditional software supply chain best practices \cite{Okafor2022SoKAOA,Ishgair2024SoKADA}.

\subsection{Pre-Deployment Mitigations}

Technical controls provide concrete mechanisms to reduce supply chain attack surface and detect compromises in Web3 systems.

\subsubsection{Dependency Verification}

Dependency verification establishes cryptographic integrity checks for all software components. Package managers should verify checksums and signatures for every downloaded dependency, with the appropriate means such as lockfiles. Also for smart contract dependencies, developers should pin specific versions and verify that imported library code matches audited versions.
Automated dependency auditing tools scan dependency trees for known vulnerabilities \cite{dirty-waters}. The npm audit command checks packages against vulnerability databases, while specialized Web3 tools like Slither can detect common smart contract vulnerability patterns in smart contract dependencies. 
Dependency review processes should verify not only vulnerability reports but also maintainer changes and unusual package updates. A sudden transfer of ownership or unexpected version bump in a critical dependency warrants investigation before integration.

Clients of upgradable smart contracts should implement strict design controls.
ERC-7936 \cite{erc-7936} standardizes an interface for proxy contracts that allows callers to explicitly select which implementation version they want to interact with. Unlike traditional proxy patterns that only expose the latest implementation, this standard enables clients to continue using previous implementations even after a malicious upgrade occurs. This mitigates a class of supply chain attacks by ensuring that a malicious upgrade to the latest implementation cannot force all dependent contracts to immediately execute compromised code; instead, clients can verify new implementations and migrate at their own pace while continuing to use audited versions.

\subsubsection{Reproducible Builds}

Reproducible builds \cite{lamb2021reproducible} ensure that source code deterministically produces identical binaries, enabling verification that deployed artifacts match audited source code.
Build systems should record all build inputs including compiler versions, dependency hashes, and environment variables to enable reproducibility verification. Verifiable cryptographic attestations of build provenance should be put in deployed artifacts. Advanced techniques such as diverse double compilation, where multiple independent build environments produce the same binary, can further increase confidence in build integrity \cite{wheeler2005countering,Rosencrantz1737190}.

Smart contract verification means publishing source code alongside deployed bytecode, enabling public audit and reducing information asymmetry. Platforms like Etherscan, Sourcify, and Blockscout allow developers to submit source code which is recompiled and matched against on-chain bytecode.
Verified contracts become transparent: users can review the actual logic before interacting, security researchers can audit for vulnerabilities, and automated tools can analyze verified source for known patterns. Verification acts as a deterrent against malicious deployments since it makes it harder for attackers to hide backdoors in code (but not impossible).

\subsection{Build \& CI/CD Servers}
\subsubsection{Sandboxed and Containerized Builds}

CI/CD pipelines should execute builds in isolated, ephemeral containers that are destroyed immediately after completion, preventing persistence of compromised environments across builds. Containerization technologies like Docker provide process isolation that limits the blast radius of compromised build dependencies: malicious code in a build dependency can access only the container's filesystem and network, not the underlying CI/CD infrastructure or other projects' artifacts. Build containers should be stateless and reconstructed from known-good images for each build, eliminating the possibility of persistent backdoors in build environments.

Secret management in containerized builds requires strict access controls and minimal exposure windows. Long-lived secrets are prohibited. Instead, secrets should be injected at runtime from dedicated secret management systems like HashiCorp Vault, AWS Secrets Manager, or the CI/CD platform's native secret store. Critically, these secrets should be provided only to the specific deployment step that requires them, not to earlier build or test stages where compromised dependencies might exfiltrate them. This principle of least privilege means build containers compile and test code without access to production credentials, while a separate, minimal deployment container receives credentials only for the transaction signing or smart contract publishing operation.

\subsubsection{Multi-Signature Deployment}

Multi-signature mechanisms distribute deployment authority across multiple independent key holders, preventing single points of compromise. Rather than a single private key controlling contract deployment or upgrades, a multi-sig scheme requires M-of-N signatures before execution.

For deployment operations, development teams should use multi-signature wallets where multiple team members must approve transactions. This prevents a compromised developer machine or stolen deployment key from unilaterally deploying malicious contracts. Time-locks can add additional security by enforcing delays between proposal and execution, allowing stakeholders to review pending operations.

Upgrade mechanisms for smart contracts should always incorporate multi-signature approval. There is reusable code for that: popular patterns like OpenZeppelin's TimelockController require multiple authorized addresses to propose, queue, and execute upgrades, with configurable delays for community review.

\subsection{Process and Policy Mitigations}

Beyond technical controls, organizational processes and policies form critical defenses against supply chain attacks in Web3 development.

\subsubsection{Security Audits}

\paragraph{Frontend audit} 
Frontend audits in Web3 applications must examine in particular the blockchain interaction layer. Web3 auditors should verify that all npm packages match their published versions, analyze transaction construction logic for parameter manipulation vulnerabilities, and review wallet integration code for key exposure risks. Critical areas include examining how transaction objects are built before being passed to wallet interfaces, and which dependencies have access to this. Frontend audits should also verify that the deployed application matches the audited source code, checking build reproducibility and deployment integrity. Given that frontend compromises can directly lead to fund theft through transaction manipulation, these audits represent a critical control point that is often overlooked in favor of smart contract audits alone.

\paragraph{Smart contract audit} 
Smart contract audits should examine both contract logic and the entire dependency chain. Professional audit firms should verify that imported libraries match their claimed versions, review all external calls to smart contract or onchain dependencies, and validate that no unexpected code modifications occurred during integration.
Pre-deployment audits should include dependency tree analysis, identifying all direct and transitive dependencies, checking each smart contract dependency against known secure versions.
Pre-deployment audits should also include reviewing maintainer history for suspicious changes.

\subsubsection{Incident Response Planning}

Web3 projects must prepare for supply chain compromise with pre-defined response procedures. Incident response plans should identify stakeholders and communication channels, define thresholds for different response levels, and establish standard operation procedures for emergency contract pausing or upgrade mechanisms.

Key compromise scenarios require specific response protocols. If deployment keys are suspected of compromise, teams should immediately rotate credentials, audit recent deployments for unauthorized changes, and communicate with users about potential risks. Smart contract upgrades should be reviewed by multiple parties even during emergency response. Good protocols are made resistent to key compromise \cite{trailofbits}.
Post-incident analysis should identify the attack vector, assess impact across deployed systems, and implement preventive controls. The transparency of blockchain enables detailed forensic analysis of attack transactions.

\subsubsection{Developer Education and Awareness}

Supply chain security requires awareness in development teams. Training programs should cover common attack vectors specific to Web3, such as malicious npm packages targeting private keys, compromised development tools modifying transaction parameters, and social engineering attacks against package maintainers.

Developers should be educated on secure credential management practices, including hardware wallet usage for deployment keys, environment variable isolation, and credential exposure in version control. Teams should establish clear policies for dependency management, such as requiring manual review before adding new dependencies, maintaining an approved package list, and documenting the purpose of each dependency.

Security awareness should extend beyond development teams to include operations staff managing CI/CD systems, community members reviewing upgrade proposals, and end users. The decentralized nature of Web3 means security responsibility is distributed across multiple stakeholders rather than concentrated in traditional security teams.

\subsection{Post-Deployment Mitigations}

\subsubsection{Monitoring of Live Web3 Systems}

Continuous monitoring and alerting systems detect supply chain threats in real-time for Web3 frontends and development infrastructure. Dependency monitoring services watch for new vulnerabilities disclosed in used packages, maintainer account compromises, or suspicious package updates. Tools like Dependabot and Renovate automate dependency updates while flagging unusual changes for manual review. Frontend monitoring should track deployment changes, and integrity of what is deployed in production. Wallet software could indicate supply chain attacks in progress, live.

On-chain monitoring focuses on detecting compromises in deployed smart contracts and their dependencies. Web3-specific tools like Tenderly and Forta enable runtime monitoring of deployed contracts, detecting anomalous transactions or unexpected state changes that might indicate exploitation of supply chain vulnerabilities. Organizations should establish automated alerts for critical events such as unexpected changes to high-privilege contracts, unusual transaction patterns from deployment addresses, or upgrades to on-chain libraries that dependent contracts rely upon. 

By combining frontend dependency monitoring with on-chain runtime surveillance, Web3 projects can detect and respond to Web3 supply chain attacks targeting either their application layer or the blockchain layer.

\subsubsection{Transaction Simulation and Analysis}

Transaction simulation represents one powerful mitigation against frontend supply chain attacks at the wallet level. Modern wallets increasingly simulate transactions before user signing to display the net effect in human-readable terms, such as "You will send 100 USDC and receive 0.05 WETH" or "You will lose your Bored Ape NFT." This approach directly counters transaction manipulation attacks by revealing the actual outcome of potentially malicious parameters, even when the raw transaction data appears complex or obfuscated.

Simulation engines execute transactions against current blockchain state in a local environment, computing the precise state changes that would result from signing. By analyzing token transfers, NFT ownership changes, and contract state mutations, simulators can detect when frontend-manipulated parameters would produce outcomes inconsistent with user intent. For example, if compromised frontend code modified a swap transaction to use an unfavorable exchange rate or changed the recipient address, simulation would reveal that the user receives far less value than expected or that assets are being sent to an unexpected address.

\section{Conclusion}

Web3 applications depend on complex software supply chains that can introduce critical vulnerabilities with catastrophic financial consequences. This paper presented a systematic analysis of supply chain security challenges unique to Web3, examining how blockchain properties create novel attack vectors beyond traditional software systems.

We presented a threat model categorizing Web3 supply chain attacks across blockchain nodes, smart contract dependencies, frontend libraries, and development tooling. Our analysis documented critical vulnerabilities including RPC endpoint manipulation, malicious library upgrades, and frontend dependency poisoning, with these attack vectors demonstrated by real incidents. We proposed a defense-in-depth strategy combining technical controls, specialized audits, and continuous monitoring.
This paper benefits the entire Web3 ecosystem, from individual developers to suppliers of dependencies used by thousands of applications, with actionable insights for DeFi protocols, wallet developers, and blockchain platforms.

As Web3 systems manage increasing value and expand into critical infrastructure, securing their software supply chains becomes an existential requirement.

\section{Acknowledgments}
I would like to thank first and foremost Raphina Liu and Monica Jin for their essential contributions to this line of research. I also thank Javier Ron, Sofia Bobadilla and Benoit Baudry for riding on-chain together. Finally, I thank all the Chains team members for the invaluable conversations on software supply chains. Stealth gratitude to Cookie Slayer.
This work was supported by the WASP program funded by Knut and Alice Wallenberg Foundation, the Swedish Foundation for Strategic Research (SSF) and the Ethereum Foundation.

\bibliographystyle{plain}
\bibliography{references}

\begin{thebibliography}{10}

\bibitem{bybit}
In-{Depth} {Technical} {Analysis} of the {Bybit} {Hack}.
\newblock
  \url{https://www.nccgroup.com/research-blog/in-depth-technical-analysis-of-the-bybit-hack/}.

\bibitem{ua-parser}
No {Unaccompanied} {Miners}: {Supply} {Chain} {Compromises} {Through} {Node}.js
  {Packages}.
\newblock
  \url{https://cloud.google.com/blog/topics/threat-intelligence/supply-chain-node-js}.

\bibitem{parity}
The {Parity} {Wallet} {Hack} {Explained} - {OpenZeppelin} blog.
\newblock
  \url{https://www.openzeppelin.com/news/on-the-parity-wallet-multisig-hack-405a8c12e8f7}.

\bibitem{badgerdao}
{TRM} {Investigates}: {BadgerDAO} {DeFi} {Protocol} {Hacked} {\textbar} {TRM}
  {Blog}.
\newblock
  \url{https://www.trmlabs.com/resources/blog/trm-investigates-badgerdao-defi-protocol-hacked}.

\bibitem{ledger_security_2023}
Security {Incident} {Report}.
\newblock \url{https://www.ledger.com/blog/security-incident-report}, 2023.

\bibitem{vibecoding}
User solana wallet exploited in first case of ai poisoning attack.
\newblock \url{https://www.bitget.com/news/detail/12560604366774}, 2024.

\bibitem{rescana_malicious_2025}
Malicious {Crypto}-{Stealing} {VSCode} {Extensions} {Target} {OpenVSX} and {AI}
  {Code} {Editors}: {Threat} {Analysis} and {Mitigation}.
\newblock
  \href{https://www.rescana.com/post/malicious-crypto-stealing-vscode-extensions-target-openvsx-and-ai-code-editors-threat-analysis-and}{link},
  2025.

\bibitem{trailofbits}
Maturing your smart contracts beyond private key risk.
\newblock
  \url{https://blog.trailofbits.com/2025/06/25/maturing-your-smart-contracts-beyond-private-key-risk/},
  2025.

\bibitem{northkorea}
North {Korean} hackers stealing record sums, researchers say.
\newblock \url{https://www.bbc.com/news/articles/cwy8z7wxe03o}, 2025.

\bibitem{Andersson1908608}
Vivi Andersson.
\newblock Geth rebuild : Verifiable builds for go ethereum.
\newblock Master's thesis, KTH, School of Electrical Engineering and Computer
  Science (EECS), 2024.

\bibitem{event-stream}
Iosif Arvanitis, Grigoris Ntousakis, Sotiris Ioannidis, and Nikos Vasilakis.
\newblock A systematic analysis of the event-stream incident.
\newblock In {\em Proceedings of the 15th European Workshop on Systems
  Security}, pages 22--28, 2022.

\bibitem{belenkov2025sok}
Nikita Belenkov, Valerian Callens, Alexandr Murashkin, Kacper Bak, Martin
  Derka, Jan Gorzny, and Sung-Shine Lee.
\newblock Sok: A review of cross-chain bridge hacks in 2023.
\newblock {\em arXiv preprint arXiv:2501.03423}, 2025.

\bibitem{leftpad}
Md~Atique~Reza Chowdhury, Rabe Abdalkareem, Emad Shihab, and Bram Adams.
\newblock On the untriviality of trivial packages: An empirical study of npm
  javascript packages.
\newblock {\em IEEE Transactions on Software Engineering}, 48(8):2695--2708,
  2021.

\bibitem{houy2023security}
Sabine Houy, Philipp Schmid, and Alexandre Bartel.
\newblock Security aspects of cryptocurrency wallets: a systematic literature
  review.
\newblock {\em ACM Computing Surveys}, 56(1):1--31, 2023.

\bibitem{huang2024overview}
Renke Huang, Jiachi Chen, Yanlin Wang, Tingting Bi, Liming Nie, and Zibin
  Zheng.
\newblock An overview of web3 technology: Infrastructure, applications, and
  popularity.
\newblock {\em Blockchain: Research and Applications}, 5(1):100173, 2024.

\bibitem{Ishgair2024SoKADA}
Eman~Abu Ishgair, Marcela~S. Melara, and Santiago Torres-Arias.
\newblock Sok: A defense-oriented evaluation of software supply chain security.
\newblock {\em ArXiv}, abs/2405.14993, 2024.

\bibitem{crystal-clear}
Monica Jin, Raphina Liu, and Martin Monperrus.
\newblock On-chain analysis of smart contract dependency risks on ethereum.
\newblock Technical Report 2503.19548, arXiv, 2025.

\bibitem{Ladisa2023JourneyTTA}
Piergiorgio Ladisa, Serena~Elisa Ponta, A.~Sabetta, Matias Martinez, and
  Olivier Barais.
\newblock Journey to the center of software supply chain attacks.
\newblock {\em IEEE Security \& Privacy}, 21:34--49, 2023.

\bibitem{lamb2021reproducible}
Chris Lamb and Stefano Zacchiroli.
\newblock Reproducible builds: Increasing the integrity of software supply
  chains.
\newblock {\em IEEE Software}, 39(2):62--70, 2021.

\bibitem{raphinathesis}
Raphina Liu.
\newblock Dirty-waters: Investigation of the software supply chain of
  javascript cryptocurrency wallets.
\newblock Master's thesis, Stockhom University, 2024.

\bibitem{dirty-waters}
Raphina Liu, Sofia Bobadilla, Benoit Baudry, and Martin Monperrus.
\newblock Dirty-waters: Detecting software supply chain smells.
\newblock In {\em Proceedings of the 33rd ACM International Conference on the
  Foundations of Software Engineering, Tool track}, 2025.

\bibitem{erc-7936}
Raphina Liu, Monica Jin, and Martin Monperrus.
\newblock {ERC}-7936: {Versioned} {Proxy} {Contract} {Interface}.
\newblock \url{https://eips.ethereum.org/EIPS/eip-7936}.

\bibitem{Ma2024DecodingWIA}
Kai Ma, Zhuo Wang, Yanjie Zhao, and Haoyu Wang.
\newblock Decoding web3: In-depth analysis of the third-party package supply
  chain.
\newblock {\em Proceedings of the 15th Asia-Pacific Symposium on Internetware},
  2024.

\bibitem{Ohm2020BackstabbersKCA}
Marc Ohm, H.~Plate, Arnold Sykosch, and M.~Meier.
\newblock Backstabber's knife collection: A review of open source software
  supply chain attacks.
\newblock {\em Detection of Intrusions and Malware, and Vulnerability
  Assessment}, 12223:23 -- 43, 2020.

\bibitem{Okafor2022SoKAOA}
C.~Okafor, Taylor~R. Schorlemmer, Santiago Torres-Arias, and James~C. Davis.
\newblock Sok: Analysis of software supply chain security by establishing
  secure design properties.
\newblock {\em Proceedings of the 2022 ACM Workshop on Software Supply Chain
  Offensive Research and Ecosystem Defenses}, 2022.

\bibitem{Rosencrantz1737190}
Niklas Rosencrantz.
\newblock Diverse double-compiling to harden cryptocurrency software.
\newblock Master's thesis, KTH, School of Electrical Engineering and Computer
  Science (EECS), 2023.

\bibitem{Sommer2023AMAA}
Jannik~Lucas Sommer, Magnus~M{\o}lgaard Lund, Nicola Cibin, and Michele Albano.
\newblock A method and platform for security advisory dissemination leveraging
  web3 technologies.
\newblock {\em 2023 IEEE International Conference on Blockchain (Blockchain)},
  pages 265--272, 2023.

\bibitem{thompson1984reflections}
Ken Thompson.
\newblock Reflections on trusting trust.
\newblock {\em Communications of the ACM}, 27(8):761--763, 1984.

\bibitem{wheeler2005countering}
David~A Wheeler.
\newblock Countering trusting trust through diverse double-compiling.
\newblock In {\em 21st Annual Computer Security Applications Conference
  (ACSAC'05)}, pages 13--pp. IEEE, 2005.

\bibitem{Williams2025ResearchDIA}
Laurie Williams, Giacomo Benedetti, Sivana Hamer, Ranindya Paramitha, Imranur
  Rahman, Mahzabin Tamanna, Greg Tystahl, Nusrat Zahan, Patrick Morrison,
  Yasemin Acar, Michel Cukier, Christian K{\"a}stner, A.~Kapravelos, Dominik
  Wermke, and William Enck.
\newblock Research directions in software supply chain security.
\newblock {\em ACM Transactions on Software Engineering and Methodology}, 34:1
  -- 38, 2025.

\bibitem{Zhu2024ESPRAEA}
Joshua Zhu, Arnav Vora, Kevin Zhao, Andy Huang, Alexander Edwards, Eddie He,
  Sanjay Krishna, Ryan Chang, Alan Wu, and Layah Vigneaud.
\newblock Espr: An ethereum-sourced package registry for software supply chain
  security.
\newblock {\em 2024 IEEE MIT Undergraduate Research Technology Conference
  (URTC)}, pages 1--5, 2024.

\end{thebibliography}

\end{document}